\documentclass[journal]{IEEEtran}
\usepackage{amsmath}
\usepackage{dblfloatfix} 
\usepackage{cite}

\ifCLASSINFOpdf
\else   %dsda
\usepackage[dvips]{graphicx}
\fi
\usepackage{url}
\usepackage{balance}
\usepackage{mathtools}
\usepackage{xcolor}
\usepackage{graphicx}
\usepackage{color}
\usepackage{soul}
\usepackage{placeins}
\usepackage{float}
\usepackage{hyperref}
\usepackage{tabularx,colortbl}
\usepackage{amssymb}
\usepackage{bm}
\usepackage{subfigure}
\usepackage{tabularx}
\usepackage{threeparttable}
\usepackage{subfigure}
\usepackage{tablefootnote}
\usepackage{amsthm}

\newcounter{MYtempeqncnt}
\begin{document}
\title{
%On the Statistics of the Bivariate $\alpha$-${\cal{F}}$ Distribution
On the Secrecy Performance of \\$\alpha$-${\cal{F}}$ Channels with Pointing Errors}

\author{

Gabriel M. C. Neves, Hugerles S. Silva, Higo T. P. Silva, Wamberto J. L. Queiroz, Felipe A. P. Figueiredo and Rausley A. A. de Souza

\thanks{
This work was partially supported by CNPq (311470/2021-1 and 403827/2021-3), by FCT - Fundação para a Ciência e Tecnologia, I.P. by project reference UIDB/50008/2020, and DOI identifier 10.54499/UIDB/50008/2020, https://doi.org/10.54499/UIDB/50008/2020, by the projects XGM-AFCCT-2024-2-5-1, XGM-FCRH-2024-2-1-1, and XGM-AFCCT-2024-9-1-1 supported by xGMobile -was partially supported by EMBRAPII-Inatel Competence Center on 5G and 6G Networks, with financial resources from the PPI IoT/Manufatura 4.0 from MCTI grant number 052/2023, signed with EMBRAPII, and by Fapemig (PPE-00124-23 and APQ-03162-24). 
G. M. C. Neves and H. S. Silva are with University of Brasília, Federal District, Brazil (e-mail: gabriel.manhaes@aluno.unb.br, hugerles.silva@unb.br). H. S. Silva is also with Instituto de Telecomunicações and Departamento de Eletrónica, Telecomunicações e Informática, Universidade de Aveiro, Campus Universitário de Santiago, 3810-193 Aveiro, Portugal  (e-mail: hugerles.silva@av.it.pt). 
%H. T. P. Silva is with the Electrical Engineering Department, Federal Institute of Paraíba (IFPB), Paraíba, Brazil. (e-mail: higo.silva@ee.ufcg.edu.br). 
W. J. L. Queiroz is with the Department of Electrical Engineering, Federal University of Campina Grande, Campina Grande, Paraíba, Brazil (e-mail: wamberto@dee.ufcg.edu.br). H. T. P. Silva, F. A. P. Figueiredo, and R. A. A. de Souza are with the National Institute of Telecommunications (Inatel), Santa Rita do Sapucaí 37540-000, Brazil. (e-mail: higo.thaian@posdoc.inatel.br, felipe.figueiredo@inatel.br, rausley@inatel.br)}
\vspace{-2em}
}

\markboth{Submitted to IEEE Wireless Communications Letters, January~2025}{}%
\maketitle

\begin{abstract}
This paper investigates the physical layer security (PLS) performance of $\boldsymbol{\alpha}$-${\cal{\boldsymbol{F}}}$ fading channels with pointing errors under passive and active eavesdropping scenarios. 
Novel analytical expressions are derived for key PLS metrics, including the probability of strictly positive secrecy capacity, the average secrecy capacity, and the secure outage probability. 
An asymptotic analysis is also investigated to provide further insights into the system behavior under high signal-to-noise ratio conditions. 
The analytical results are validated through Monte Carlo simulations, with several performance curves presented for a range of channel and system parameters. 
All expressions derived in this work are original and have not been previously published.
\end{abstract}


\IEEEpeerreviewmaketitle
\section{Introduction}\label{sec1}
%%%
\IEEEPARstart{P}{hysical} layer security~(PLS) has gained significant attention from researchers over the recent years due to its ability to prevent eavesdropping without relying on data encryption at higher layers of the communication protocol~\cite{Shiu}. 
This approach is particularly relevant in digital communications over fading channels, where the inherent characteristics of the physical layer can be exploited to enhance secrecy. 
As a result, there is a growing interest in investigating and optimizing the performance of PLS techniques in these environments.

Research findings concerning PLS are evaluated under fading channels and/or pointing errors in the literature.
In~\cite{Lei}, the secrecy performance of the classic Wyner’s wiretap model under generalized-$K$ fading channels is studied where closed-form expressions for the probability of strictly positive secrecy capacity~(SPSC), the average secrecy capacity~(ASC), and the secure outage probability~(SOP) are derived.
In~\cite{Bhargav}, expressions for the probability of SPSC and a lower bound for the SOP under $\kappa$-$\mu$ fading are presented.
The secrecy capacity under Fisher-Snedecor $\mathcal{F}$ fading channels is evaluated in~\cite{BadarnehFisher}, where exact closed-form expressions are derived for the probability of SPSC and ASC alongside an asymptotic analysis for the ASC.
In~\cite{Moualeu}, the PLS of wireless systems under $\alpha$-$\kappa$-$\mu$ and $\alpha$-$\eta$-$\mu$ fading is analyzed employing asymptotic and bound expressions for the SOP.
A secrecy capacity analysis is performed in~\cite{Lei2017} and~\cite{Mathur} over independent and correlated $\alpha$-$\mu$ fading channels, respectively.
The secrecy performance of a free-space optical system is presented in~\cite{Han} using different detection techniques, considering Fisher-Snedecor $\mathcal{F}$ channels with pointing error. 
The authors also derive expressions for the ASC, SOP, and the probability of SPSC.
In~\cite{Tota}, a secrecy capacity study is made over a reconfigurable intelligent surface-assisted terahertz (THz) system with pointing errors.

This paper evaluates the secrecy performance of $\alpha$-${\cal{F}}$ fading channels with pointing errors using the probability of SPSC, ASC, and SOP metrics.
To the best of the authors’ knowledge, this is the first work where the PLS performance is investigated in this context.
We adopt the $\alpha$-$\cal{F}$ distribution with pointing errors since it is simple and jointly characterizes the beam misalignment, the small- and large-scale fading, and the nonlinearity of the communication channel.
The $\alpha$-$\cal{F}$ with pointing error distribution is also a promising model to be adopted in THz channels and suitable for applications in wireless emerging scenarios~\cite{Almeida2023}. 
All the expressions derived in this work are unprecedented in the literature. 

In a nutshell, the main contributions of this article are: (i) new and exact expressions are deduced for the probability of SPSC and ASC over $\alpha$-$\cal{F}$ fading channels with pointing errors; (ii) a lower bound for the SOP is presented, and (iii) an asymptotic analysis is performed to provide insights into the channel's effect on secrecy performance.    

% The remainder of the paper is organized as follows. 
% Section~\ref{SystemModel} describes the system and channel models. 
% Some important secrecy performance metrics\textemdash the SPSC, and the exact and the asymptotic ASC\textemdash are presented in Section~\ref{PerformanceAnalysis}. 
% Section~\ref{OutageCapacity} presents a lower bound for the SOP.
% Section~\ref{results} shows the numerical results.
% Section~\ref{conclusao} brings the conclusions of the paper.

\section{System and Channel Models}\label{SystemModel}
\begin{figure}
    \centering
    \includegraphics[width=0.95\linewidth]{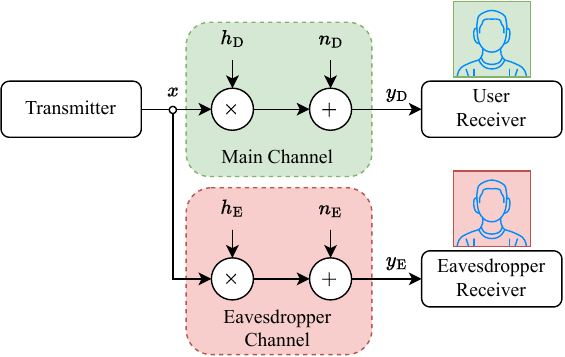}
    \caption{System model illustration, in which an eavesdropper intercepts a legitimate user's communication.}
    \label{fig:wiretapSystem}
\end{figure}

Let us assume a three-node classic model in which a transmitter communicates with a legitimate user receiver while an eavesdropper can intercept the directive beam of the transmitter's antenna, as shown in Fig. \ref{fig:wiretapSystem}. 
It is assumed that the transmitter, the user, and the eavesdropper are equipped with highly directive antennas, making them susceptible to pointing errors. 
This scenario is particularly relevant for THz band communications, where all link components employ directive antennas to counteract significant path losses. 
%Additionally, it is assumed that the eavesdropper can intercept the directive beam of the transmitter's antenna. 
In these conditions, it is valid to consider that the channels are characterized by $\alpha$-${\cal{F}}$ fading with pointing errors.
The passive and active eavesdropping scenarios are analyzed.
%In this study, we analyze the scenarios of passive and active eavesdropping. 
In the former, only the main link's channel state information (CSI) is known at the transmitter. 
In the latter, the CSI of all the links is available at the transmitter.

The signal at the receiver's matched filter output is given by $y_i = h_{\text{l}_{i}}h_ix+n_i$, in which $i\in \{{\text{D}},{\text{E}}\}$, with ${\text{D}}$ and ${\text{E}}$ respectively representing the indicators corresponding to the legitimate user and the eavesdropper, $h_{\text{l}_i}$ is the path loss, $h_i$ is the complex channel gain, $x$ represents the transmitted signal, and $n_i$ is the additive white Gaussian noise with zero mean and unit variance.
For THz systems, $h_{\text{l}_{i}}=h_{\text{fl}_{i}}h_{\text{al}_{i}}$, where $h_{\text{fl}_{i}} = {c \sqrt{G_t G_{r_i}}}/\left({4\pi f d_{i}}\right)$ models the propagation gain, in which $G_t$ and $G_{r_{i}}$ are, respectively, the gains of the transmit- and receive-antenna, $c$ is the speed of light, $f$ is the operating frequency, and $d_{i}$ is the distance between the transmitter and the receiver. 
The molecular absorption gain is characterized $h_{\text{al}_{i}} = {\exp}\left(-\kappa(f)d_{i}/2\right)$, where $\kappa(f)$ denotes the absorption coefficient.
More details about $h_{\text{l}_{i}}$ can be found in~\cite{Boulogeorgos}. 

For the fading model adopted, the probability density
function (PDF) and cumulative distribution function (CDF) of the instantaneous signal-to-noise ratio~(SNR) $\Gamma_i$ are respectively written as \cite{Almeida2023}
    \begin{IEEEeqnarray}{lcl}\label{eq:PDFSNR}
        f_{\Gamma_{i}}(\gamma_{i}) = \frac{\Lambda_{i}}{2\gamma_{i}} {\rm H}^{2,1}_{2,2}\left[\Theta_i\gamma_{i}^{\frac{\alpha_i}{2}}\bigg|
        \begin{array}{c}
            (1-m_{i},1) ,(z_{i}^{2}/\alpha_{i}+1,1) \\
            (\mu_{i},1) ,(z_{i}^{2}/\alpha_{i},1)
        \end{array}
        \right]\IEEEeqnarraynumspace
    \end{IEEEeqnarray}
and
    \begin{IEEEeqnarray}{lcl}\label{eq:CDFSNR}
        &&F_{\Gamma_i}(\gamma_{i}) = \frac{\Lambda_{i}}{\alpha_{i}}\nonumber\\
        &&\times{\rm H}^{2,2}_{3,3}\bigg[\Theta_i\gamma_{i}^{\frac{\alpha_i}{2}}\bigg|
        \begin{array}{c}
            (1-m_{i},1), (1,1), (z_{i}^{2}/\alpha_{i}+1,1) \\
            (\mu_{i},1), (z_{i}^{2}/\alpha_{i},1), (0,1)
        \end{array}
        \bigg],\IEEEeqnarraynumspace
    \end{IEEEeqnarray}
in which $\Lambda_{i} = z_{i}^2/[\Gamma(\mu_{i})\Gamma(m_{i})]$, $\Theta_i = \frac{\mu_{i}}{m_{i} - 1} \left[  \frac{z_{i} \sqrt{\lambda_{i}}}{\sqrt{ \bar{\gamma}_{i}( z_{i}^2 + 2 ) }}\right]^{\alpha_{i}}$,
$\bar{\gamma}_{i}$ is the average SNR, $\alpha_i>0$ characterizes the nonlinearity of the propagation medium, $\mu_i>0$ represents the number of multipath clusters, $m_i>1$ is the shadowing parameter, $z_{i}>0$ is the pointing error intensity, $\lambda_{i}$ is given by~\cite[Eq. (4)]{Badarneh}, H$[\cdot]$ is the Fox H-function~\cite[Eq. (1.2)]{Maithai}, and $\Gamma(\cdot)$ is the Gamma function. %~\cite[Eq. (8.310)]{Gradshteyn}.
For $z_{i}\xrightarrow{}\infty$, the case of the non-pointing error is found. 
From \eqref{eq:PDFSNR} and \eqref{eq:CDFSNR}, the statistics for the $\alpha$-$\mu$ ($m_{i}\rightarrow \infty)$ and the Fisher-Snedecor $\cal{F}$ ($\alpha_{i}=2$) distributions with pointing errors and their inclusive ones can be obtained.

\section{Secrecy Capacity Analysis}\label{PerformanceAnalysis}

Under active eavesdropping, the system's instantaneous secrecy capacity is defined as~\cite[Eq. (11)]{Bhargav}
    \begin{equation}\label{eq:InstCapacity}
        C_\text{s}(\gamma_{\text{D}},\gamma_{\text{E}}) = [\log_2(1+\gamma_{\text{D}})-\log_2(1+\gamma_{\text{E}})]^{+},
    \end{equation}
in which $\log_2(1+\gamma_{\text{D}})$ and $\log_2(1+\gamma_{\text{E}})$ are the capacities corresponding to the user and eavesdropper channels, respectively, and $[x]^{+} = \max\{x, 0\}$.

\subsection{Strictly Positive Secrecy Capacity}

The probability of SPSC is defined as the probability of the existence of a secrecy capability, that is, the probability that the capacity of the legitimate channel exceeds the capacity of the eavesdropping channel. 
Mathematically~\cite[Eq. (10)]{BadarnehFisher}, 
    \begin{IEEEeqnarray}{lcl}\label{eq:CapacidadeAsym}
      \mathbb{P}_{0}=\Pr\{C_\text{s}(\gamma_{\text{D}},\gamma_{\text{E}})>0\}=\int_{0}^{\infty}f_{\Gamma_{\text{D}}}(\gamma_{\text{D}})F_{\Gamma_{\text{E}}}(\gamma_{\text{D}})\text{d}\gamma_{\text{D}}.\IEEEeqnarraynumspace
    \end{IEEEeqnarray}
Substituting \eqref{eq:PDFSNR} and \eqref{eq:CDFSNR} into \eqref{eq:CapacidadeAsym}, then making the variable change $u=\gamma_{\text{D}}^{\alpha_{\text{E}}/2}$, using~\cite[Eq. (2.8.4)]{HTransforms}, and proceeding with simplifications, it is possible to express the probability of SPSC under $\alpha$-${\cal{F}}$ fading channels with pointing errors as in~\eqref{eq:ExistCapacidade}.

\begin{figure*}%[!h]
    \begin{align}\label{eq:ExistCapacidade}
        \mathbb{P}_{0} = \frac{\Lambda_{\text{D}}\Lambda_{\text{E}}}{\alpha_{\text{E}}^2}{\rm H}^{4,3}_{5,5}\left[\frac{\Theta_\text{D}}{\Theta_{\text{E}}^{\frac{\alpha_\text{D}}{\alpha_\text{E}}}}\bigg|
            \begin{array}{c}
                (1-m_\text{D},1), (1-\mu_\text{E},\frac{\alpha_\text{D}}{\alpha_\text{E}}), (1-\frac{z_{\text{E}}^{2}}{\alpha_\text{E}},\frac{\alpha_\text{D}}{\alpha_\text{E}}),(1,\frac{\alpha_\text{D}}{\alpha_\text{E}}), (\frac{z_\text{D}^{2}}{\alpha_\text{D}}+1,1)\\
                (\mu_\text{D},1), (\frac{z_{\text{D}}^{2}}{\alpha_\text{D}},1), (m_\text{E},\frac{\alpha_\text{D}}{\alpha_\text{E}}),(0,\frac{\alpha_\text{D}}{\alpha_\text{E}}), (-\frac{z_{\text{E}}^{2}}{\alpha_\text{E}},\frac{\alpha_\text{D}}{\alpha_\text{E}})
        \end{array}
            \right].
    \end{align}
\hrulefill
\end{figure*}

\subsection{Average Secrecy Capacity}

The ASC, defined as the maximum achievable secrecy rate, can be derived from~(\ref{eq:InstCapacity}) as~\cite[Eq. (14)]{BadarnehFisher}
    \begin{align}\label{eq:AveraCapacity}
        \bar{C}_{\text{s}} &= \int_{0}^{\infty}\int_{0}^{\infty}C_\text{s}(\gamma_{\text{D}},\gamma_{\text{E}}) f_{\Gamma_{\text{D}}}(\gamma_{\text{D}})f_{\Gamma_{\text{E}}}(\gamma_{\text{E}})\text{d}\gamma_{\text{D}}\text{d}\gamma_{\text{E}}\nonumber\\
        &=\frac{I_1+I_2-I_3}{\ln(2)},
    \end{align}
in which $I_1 = \int_{0}^{\infty}\ln(1+\gamma_{\text{D}})f_{\Gamma_{\text{D}}}(\gamma_{\text{D}})F_{\Gamma_{\text{E}}}(\gamma_{\text{D}})\text{d}\gamma_{\text{D}}$, $I_2 = \int_{0}^{\infty}\ln(1+\gamma_{\text{E}})f_{\Gamma_{\text{E}}}(\gamma_{\text{E}})F_{\Gamma_{\text{D}}}(\gamma_{\text{E}})\text{d}\gamma_{\text{E}}$, $I_3 = \int_{0}^{\infty}\ln(1+\gamma_{\text{E}})f_{\Gamma_{\text{E}}}(\gamma_{\text{E}})\text{d}\gamma_{\text{E}}$, since $\gamma_{\text{D}}$ and $\gamma_{\text{E}}$ are independent.
Note that $I_{1}$ and $I_{2}$ are symmetric with respect to the channel parameters. 
Therefore, these integrals can be generically represented by $\bar{I}(\mathcal{P}_{\text{x}},\mathcal{P}_{\text{y}})$, where $\mathcal{P}_{\text{x}} = \{z_{\text{x}},\alpha_\text{x},\mu_{\text{x}},z_{\text{x}},\Theta_{\text{x}}\}$ and $\mathcal{P}_{\text{y}} = \{z_{\text{y}},\alpha_{\text{y}},\mu_{\text{y}},z_{\text{y}},\Theta_{\text{y}}\}$ denote the sets of channel parameters. 
Applying~(\ref{eq:PDFSNR}) and~(\ref{eq:CDFSNR}) into $\bar{I}(\mathcal{P}_{\text{x}},\mathcal{P}_{\text{y}})$ and using~\cite[id. 01.04.26.0003.01]{Wolfram}, the general integral $\bar{I}(\mathcal{P}_{\text{x}},\mathcal{P}_{\text{y}})$ can be written as shown in~(\ref{eq:I1Um}).
Then, by writing the first two Fox H-functions of~(\ref{eq:I1Um}) in terms of the Mellin-Barnes integral~\cite[Eq. (1.2)]{Maithai}, it is possible to obtain $\bar{I}(\mathcal{P}_{\text{x}},\mathcal{P}_{\text{y}})$ as in~(\ref{eq:I1Dois}). 
Solving the integral $\mathcal{I}$ by means of~\cite[Eq. (2.8)]{Maithai} and using~\cite[Eq. (28)]{Alhennawi} in sequence, it is possible to derive a closed-form solution for $\bar{I}(\mathcal{P}_{\text{x}},\mathcal{P}_{\text{y}})$ as presented in~(\ref{eq:I1Tres}) where the tuples are given by~(\ref{eq:AJ}). 
\begin{figure*}%[!t]
    \begin{align}\label{eq:I1Um}
        &\bar{I}(\mathcal{P}_{\text{x}},\mathcal{P}_{\text{y}}) =\nonumber\\ &\frac{\Lambda_{\text{x}}\Lambda_{\text{y}}}{2\alpha_{\text{y}}}\int_{0}^{\infty}\gamma_{\text{x}}^{-1}{\rm H}^{2,1}_{2,2}\left[\frac{\Theta_\text{x}}{\gamma_{\text{x}}^{-\frac{\alpha_\text{x}}{2}}}\bigg|
    \begin{array}{c}
        (1-m_{\text{x}},1), (\frac{z_{\text{x}}^{2}}{\alpha_{\text{x}}}+1,1) \\
        (\mu_{\text{x}},1), (\frac{z_{\text{x}}^{2}}{\alpha_{\text{x}}},1)
    \end{array}
    \right]{\rm H}^{2,2}_{3,3}\left[\frac{\Theta_\text{y}}{\gamma_{\text{x}}^{-\frac{\alpha_\text{y}}{2}}}\bigg|
    \begin{array}{c}
        (1-m_{\text{y}},1), (1,1), (\frac{z_{\text{y}}^{2}}{\alpha_{\text{y}}}+1,1) \\
        (\mu_{\text{y}},1), (\frac{z_{\text{y}}^{2}}{\alpha_{\text{y}}},1), (0,1)
    \end{array}
    \right]
        {\rm H}^{1,2}_{2,2}\bigg[\gamma_{\text{x}}\bigg|
        \begin{array}{c}
                (1,1),(1,1) \\
                (1,1),(0,1)
        \end{array}
         \bigg]
        \text{d}\gamma_\text{x}.
    \end{align}
\hrulefill
\begin{align}\label{eq:I1Dois}
        \bar{I}(\mathcal{P}_{\text{x}},\mathcal{P}_{\text{y}}) &= \frac{1}{(2\pi i)^2}\int_{\mathcal{L}_1}\int_{\mathcal{L}_2}\frac{\Gamma(\mu_{\text{x}}+s_1)\Gamma(z_{\text{x}}^{2}/\alpha_{\text{x}}+s_1)\Gamma(m_\text{x}-s_1)}{\Gamma(z_{\text{x}}^{2}/\alpha_{\text{x}}+1+s_1)}\Theta_{\text{x}}^{-s_1}\frac{\Gamma(\mu_\text{y}+s_2)\Gamma(z_{\text{y}}^{2}/\alpha_{\text{y}}+s_2)\Gamma(m_\text{y}-s_2)\Gamma(-s_2)}{\Gamma(z_{\text{y}}^{2}/\alpha_{\text{y}}+1+s_2)\Gamma(1-s_2)}\Theta_{\text{y}}^{-s_2}\nonumber\\
        &\times     \frac{\Lambda_{\text{x}}\Lambda_{\text{y}}}{2\alpha_{\text{y}}}    \underbrace{\int_{0}^{\infty}\gamma_{\text{x}}^{-\frac{\alpha_\text{x} s_1}{2}-\frac{\alpha_\text{y} s_2}{2}-1}{\rm H}^{1,2}_{2,2}\bigg[\gamma_{\text{x}}\bigg|
        \begin{array}{c}
                (1,1),(1,1) \\
                (1,1),(0,1)
        \end{array}
         \bigg]\text{d}\gamma_\text{x}}_{\mathcal{I}}\text{d}s_1\text{d}s_2.
    \end{align}
\hrulefill
    \begin{align}\label{eq:I1Tres}
         \bar{I}(\mathcal{P}_{\text{x}},\mathcal{P}_{\text{y}}) &= \frac{\Lambda_{\text{x}}\Lambda_{\text{y}}}{2\alpha_{\text{y}}}{\rm H}_{4,0:2,2;3,3}^{0,3:2,1;2,2}
    \left[ % args H-function
        \begin{array}{c}
            \Theta_\text{x} \\
            \Theta_\text{y}
        \end{array}
        \begin{array}{|c}(a_{j};\tilde{\alpha}_{j}^{(1)},\tilde{\alpha}_{j}^{(2)})_{j=1:4} \\
            -
        \end{array}
        \begin{array}{|c}
        (1-m_{\text{x}},1), (\frac{z_{\text{x}}^{2}}{\alpha_{\text{x}}}+1,1) \\
        (\mu_{\text{x}},1), (\frac{z_{\text{x}}^{2}}{\alpha_{\text{x}}},1)
        \end{array}
        \begin{array}{|c}
        (1-m_{\text{y}},1), (1,1), (\frac{z_{\text{y}}^{2}}{\alpha_{\text{y}}}+1,1) \\
        (\mu_{\text{y}},1), (\frac{z_{\text{y}}^{2}}{\alpha_{\text{y}}},1), (0,1)
        \end{array}
    \right].
    \end{align}
\hrulefill
\begin{align}\label{eq:AJ}
(a_{j};\tilde{\alpha}_{j}^{(1)},\tilde{\alpha}_{j}^{(2)})_{j=1:4} = \left(0;\frac{\alpha_\text{x}}{2},\frac{\alpha_\text{y}}{2}\right),  \left(1;-\frac{\alpha_\text{x}}{2},-\frac{\alpha_\text{y}}{2}\right),\left(1;-\frac{\alpha_\text{x}}{2},-\frac{\alpha_\text{y}}{2}\right),\left(1;\frac{\alpha_\text{x}}{2},\frac{\alpha_\text{y}}{2}\right).
\end{align}
\hrulefill
\end{figure*}
Finally, we find 
\begin{equation}\label{eq:I1PdPe}
I_{1} = \bar{I}(\mathcal{P}_{\text{D}},\mathcal{P}_{\text{E}})
\end{equation}
and
\begin{equation}\label{eq:I2PdPe}
I_{2} = \bar{I}(\mathcal{P}_{\text{E}},\mathcal{P}_{\text{D}})
\end{equation}
In turn, $I_3$ can be solved using~\cite[Eq. (2.8.4)]{HTransforms}, resulting in
\begin{equation}
    \label{eq:I3Um}
    \begin{aligned}
    &I_3 = \frac{\Lambda_{\text{E}}}{2} \\ 
    &\times{\rm H}^{4,2}_{4,4}\left[\Theta_\text{E}\bigg|
    \begin{array}{c}
        (1-m_{\text{E}},1), (0,\frac{\alpha_{\text{E}}}{2}), (1,\frac{\alpha_{\text{E}}}{2}), (\frac{z_{\text{E}}^{2}}{\alpha_{\text{E}}}+1,1) \\
        (\mu_{\text{E}},1), (\frac{z_{\text{E}}^{2}}{\alpha_{\text{E}}},1), (0,\frac{\alpha_{\text{E}}}{2}), (0,\frac{\alpha_{\text{E}}}{2})
    \end{array}
    \right].
    \end{aligned}
\end{equation}
Replacing~(\ref{eq:I1PdPe}),~(\ref{eq:I2PdPe}), and~(\ref{eq:I3Um}) in~(\ref{eq:AveraCapacity}), a new expression for the ASC under $\alpha$-${\cal{F}}$ fading channels with pointing errors is found.

\subsection{Asymptotic Secrecy Capacity}

The asymptotic secrecy capacity, defined as the ASC in high-SNR regime and denoted as $\bar{C}_\text{s}^{\infty}$, is given by~\cite[Eq. (6)]{Lei2017}
\begin{IEEEeqnarray}{lcl}\label{eq:AsympCapacity}
	    \bar{C}_{\text{s}}^{\infty} = \int_{0}^{\infty}f_{\Gamma_\text{E}}(\gamma_\text{E})\text{d}\gamma_\text{E}\int_{\gamma_\text{E}}^{\infty}\ln\left(\frac{1+\gamma_\text{D}}{1+\gamma_\text{E}}\right)f_{\Gamma_\text{D}}(\gamma_\text{D})\text{d}\gamma_\text{D}.\IEEEeqnarraynumspace
\end{IEEEeqnarray}

Making $\bar{\gamma}_\text{D} = \bar{\gamma}_\text{E} = \bar{\gamma}\rightarrow \infty$, and applying the change of variables $u = \gamma_\text{D}/\bar{\gamma}$ and $v = \gamma_\text{E}/\bar{\gamma}$, the expression in~\eqref{eq:Ascassintotica1} is obtained with
%$\phi_i = \frac{\mu_{i}}{m_{i} - 1}\left[  \frac{z_{i} \sqrt{\lambda_{i}}}{\sqrt{ ( z_{i}^2 + 2 ) }}\right]^{\alpha_{i}}$ 
$\phi_i = \Theta_i\bar{\gamma}_{i}^{\alpha_{i}/2}$  and $\rho = \bar{\gamma}_\text{D}/\bar{\gamma}_\text{E}$.
% $\phi_i = \left[  \frac{\mu_{i}^{\frac{1}{\alpha_i}}z_{i} \sqrt{\lambda_{i}}}{(m_{i} - 1)^{\frac{1}{\alpha_i}}\sqrt{ ( z_{i}^2 + 2 ) }}\right]^{\alpha_{i}}$ and $\rho = \bar{\gamma}_\text{D}/\bar{\gamma}_\text{E}$.
Expressing the Fox H-functions in terms of the Mellin-Barnes integral for the inner integrals reveals that they are power-type and can be properly solved. 
Rearranging the results regarding the Fox H-functions, the outer integrals can be solved using the Gauss-Laguerre quadrature. 
After simplifications, the asymptotic ASC is expressed as
%%%%%
\begin{figure*}%[!t]
  \begin{align}\label{eq:Ascassintotica1}
&\bar{C}_{\text{s}}^{\infty} = \frac{\Lambda_{\text{D}}\Lambda_{\text{E}}}{4} \times \\
&\Bigg[\ln(\rho)\int_{0}^{\infty}\int_{\frac{y}{\rho}}^{\infty}(xy)^{-1}{\rm H}^{2,1}_{2,2}\left[\phi_\text{E} y^{\frac{\alpha_\text{E} }{2}}\bigg|
    \begin{array}{c}
        (1-m_{\text{E} },1) ,(z_{\text{E} }^{2}/\alpha_{\text{E} }+1,1) \\
        (\mu_{\text{E} },1) ,(z_{\text{E} }^{2}/\alpha_{\text{E} },1)
    \end{array}
    \right]{\rm H}^{2,1}_{2,2}\left[\phi_\text{D}  x^{\frac{\alpha_\text{D} }{2}}\bigg|
    \begin{array}{c}
        (1-m_{\text{D}},1) ,(z_{\text{D}}^{2}/\alpha_{\text{D}}+1,1) \\
        (\mu_{\text{D}},1) ,(z_{\text{D}}^{2}/\alpha_{\text{D}},1)
    \end{array}
    \right]\text{d}x\text{d}y\nonumber\\
    &+ \int_{0}^{\infty}\int_{\frac{y}{\rho}}^{\infty}(xy)^{-1}\ln(x){\rm H}^{2,1}_{2,2}\left[\phi_\text{D} x^{\frac{\alpha_\text{D}}{2}}\bigg|
    \begin{array}{c}
        (1-m_{\text{D}},1) ,(z_{\text{D}}^{2}/\alpha_{\text{D}}+1,1) \\
        (\mu_{\text{D}},1) ,(z_{\text{D}}^{2}/\alpha_{\text{D}},1)
    \end{array}
    \right]\nonumber {\rm H}^{2,1}_{2,2}\left[\phi_\text{E} y^{\frac{\alpha_\text{E}}{2}}\bigg|
    \begin{array}{c}
        (1-m_{\text{E}},1) ,(z_{\text{E}}^{2}/\alpha_{\text{E}}+1,1) \\
        (\mu_{\text{E}},1) ,(z_{\text{E}}^{2}/\alpha_{\text{E}},1)
    \end{array}
    \right]\text{d}x\text{d}y\nonumber\\
    &- \int_{0}^{\infty}\int_{\frac{y}{\rho}}^{\infty}(xy)^{-1}\ln(y){\rm H}^{2,1}_{2,2}\left[\phi_\text{D} x^{\frac{\alpha_\text{D}}{2}}\bigg|
    \begin{array}{c}
        (1-m_{\text{D}},1) ,(z_{\text{D}}^{2}/\alpha_{\text{D}}+1,1) \\
        (\mu_{\text{D}},1) ,(z_{\text{D}}^{2}/\alpha_{\text{D}},1)
    \end{array}
    \right]\nonumber{\rm H}^{2,1}_{2,2}\left[\phi_\text{E} y^{\frac{\alpha_\text{E}}{2}}\bigg|
    \begin{array}{c}
        (1-m_{\text{E}},1) ,(z_{\text{E}}^{2}/\alpha_{\text{E}}+1,1) \\
        (\mu_{\text{E}},1) ,(z_{\text{E}}^{2}/\alpha_{\text{E}},1)
    \end{array}
    \right] \text{d}x\text{d}y\Bigg].
\end{align}  
\hrulefill%\vspace{-0.8cm}
\end{figure*}
%%%%%%
\begin{figure*}%[!t]
\setcounter{MYtempeqncnt}{\value{equation}}
\setcounter{equation}{17}
    \begin{align}\label{eq:SOPLOWEREXP}
        \text{SOP}_{\mathcal{L}} = \frac{\Lambda_\text{D}\Lambda_\text{E}}{\alpha_\text{D}\alpha_\text{E}}
        {\rm H}^{3,4}_{5,5}\left[\frac{\Theta_\text{D} e^{\frac{\alpha_\text{D} R_\text{s}}{2}}}{\Theta_\text{E}^{\frac{\alpha_\text{D}}{\alpha_\text{E}}}}\right|
            \begin{array}{c}
                (1-m_\text{D},1), (1,1), (1-\mu_\text{E},\frac{\alpha_\text{D}}{\alpha_\text{E}}), (1-\frac{z_\text{E}^{2}}{\alpha_\text{E}},\frac{\alpha_\text{D}}{\alpha_\text{E}}), (\frac{z_\text{D}^{2}}{\alpha_\text{D}}+1,1)\\
                (\mu_\text{D},1), (\frac{z_\text{D}^{2}}{\alpha_\text{D}},1), (m_\text{E},\frac{\alpha_\text{D}}{\alpha_\text{E}}), (-\frac{z_\text{E}^{2}}{\alpha_\text{E}},\frac{\alpha_\text{D}}{\alpha_\text{E}}), (0, 1)
        \end{array}
            \bigg].
    \end{align}
\setcounter{equation}{\value{MYtempeqncnt}}
\addtocounter{equation}{1}
\hrulefill
\end{figure*}
    \begin{IEEEeqnarray}{lcl}\label{eq:AsympCapacity2}
        \bar{C}_{\text{s}}^{\infty} = \frac{\Lambda_\text{D} \Lambda_\text{E}}{2} \sum_{k=1}^{N}\left(\frac{1}{\alpha_\text{E}}\varphi_1(x_k)-\frac{1}{\alpha_\text{D}}\varphi_2(x_k)\right)w_k,\IEEEeqnarraynumspace
    \end{IEEEeqnarray}
in which $x_{k}$ and $w_{k}$ denote the $N$ order Laguerre polynomial root
and quadrature weight, respectively.
Furthermore, the functions $\varphi_1(x)$ and $\varphi_2(x)$ are given by
\begin{align*}
    &\varphi_1(x) = e^{x}x^{-1}\ln(x)\nonumber\\
&\times {\rm H}^{2,1}_{2,2}\left[\phi_\text{D} x^{\frac{\alpha_\text{D}}{2}}\bigg|
    \begin{array}{c}
        (1-m_{\text{D}},1) ,(z_{\text{D}}^{2}/\alpha_{\text{D}}+1,1) \\
        (\mu_{\text{D}},1) ,(z_{\text{D}}^{2}/\alpha_{\text{D}},1)
    \end{array}
    \right]\nonumber\\
    &\times {\rm H}^{2,2}_{3,3}\left[\phi_\text{E} (\rho x)^{\frac{\alpha_\text{E}}{2}}\bigg|
    \begin{array}{c}
        (1-m_{\text{E}},1) , (1,1), (z_{\text{E}}^{2}/\alpha_{\text{E}}+1,1) \\
        (\mu_{\text{E}},1) ,(z_{\text{E}}^{2}/\alpha_{\text{E}},1), (0,1)
    \end{array}
    \right]
\end{align*}
and
\begin{align*}
    &\varphi_2(x) = e^{x}x^{-1}\ln(x/\rho)\nonumber\\
&\times {\rm H}^{2,1}_{2,2}\left[\phi_\text{E} x^{\frac{\alpha_\text{E}}{2}}\bigg|
    \begin{array}{c}
        (1-m_{\text{E}},1) ,(z_{\text{E}}^{2}/\alpha_{\text{E}}+1,1) \\
        (\mu_{\text{E}},1) ,(z_{\text{E}}^{2}/\alpha_{\text{E}},1)
    \end{array}
    \right]\nonumber\\
    &\times {\rm H}^{3,1}_{3,3}\left[\phi_\text{D} \left(\frac{x}{\rho}\right)^{\frac{\alpha_\text{D}}{2}}\bigg|
    \begin{array}{c}
        (1-m_{\text{D}},1) , (z_{\text{D}}^{2}/\alpha_{\text{D}}+1,1), (1,1) \\
        (\mu_{\text{D}},1) ,(z_{\text{D}}^{2}/\alpha_{\text{D}},1), (0,1)
    \end{array}
    \right].
\end{align*}

\section{Secrecy Outage Probability Analysis}\label{OutageCapacity}

Under passive eavesdropper, the SOP is defined as the probability that the instantaneous secrecy capacity falls below a specific target secrecy rate $R_\text{s}$, which can be written as~\cite[Eq. (5)]{Moualeu}.
The calculation of \cite[Eq. (5)]{Moualeu} is very complex due to the forms of  \eqref{eq:PDFSNR} and \eqref{eq:CDFSNR}. 
Thus, we present a lower bound for the SOP, such as~\cite[Eq. (6)]{Moualeu}
    \begin{IEEEeqnarray}{lcl}\label{eq:LowerSOP}
        \text{SOP}_{\mathcal{L}} &= \Pr[\gamma_\text{D} < e^{R_\text{s}}\gamma_\text{E}]{=}\int_{0}^{\infty}F_{\Gamma_\text{D}}(e^{R_\text{s}}\gamma_\text{E})f_{\Gamma_\text{E}}(\gamma_\text{E})\text{d}\gamma_{\text{E}}.\IEEEeqnarraynumspace
    \end{IEEEeqnarray}
Substituting (\ref{eq:PDFSNR}) and~(\ref{eq:CDFSNR}) into (\ref{eq:LowerSOP}) and using~\cite[Eq. (2.8.4)]{HTransforms}, $\text{SOP}_{\mathcal{L}}$ can be written as shown in~(\ref{eq:SOPLOWEREXP}).

\subsection{Asymptotic Secrecy Outage Probability}

Considering the SOP lower bound expressed in~\eqref{eq:SOPLOWEREXP} and following the approach presented in~\cite[Theorem 1.11]{HTransforms}, it is possible to derive the asymptotic SOP for high-SNR regime of the legitimate user. 
This metric is expressed as
\begin{equation}
    \label{eq:SOPLOWEREXP_Asymptotic}
    \text{SOP}_{\mathcal{L}}^{\infty} =   
        \frac{\Lambda_{\text{D}}\Lambda_{\text{E}}\Phi}{\alpha_\text{D}\alpha_\text{E}} \left(\frac{\Theta_\text{D} e^{\frac{\alpha_\text{D} R_\text{s}}{2}}}{\Theta_{\text{E}}^{\frac{\alpha_\text{D}}{\alpha_\text{E}}}}\right)^{\Xi},
\end{equation}
where $\Xi = \min\left(\mu_{\text{D}}, z_{\text{D}}^{2}/\alpha_\text{D},  m_\text{E}\alpha_\text{E}/\alpha_\text{D} \right)$ and (i)  for $\mu_{\text{D}}<\min
\{z_{\text{D}}^{2}/\alpha_{\text{D}},m_{\text{E}}\alpha_{\text{E}}/\alpha_{\text{D}}\}$,  
    % \begin{align}
    %     \Phi &= \frac{\Gamma(\frac{z_{\text{D}}^{2}}{\alpha_\text{D}}-\mu_\text{D})\Gamma(m_\text{E}-\frac{\mu_\text{D}\alpha_\text{D}}{\alpha_\text{E}})\Gamma(m_\text{D}+\mu_\text{D})\Gamma(\mu_\text{D})}{\Gamma(\frac{z_{\text{D}}^{2}}{\alpha_\text{D}}+1-\mu_\text{D})\Gamma(1+\frac{z_{\text{E}}^{2}}{\alpha_\text{E}}+\frac{\mu_\text{D}\alpha_\text{D}}{\alpha_\text{E}})}\nonumber\\
    %     &\times \frac{\Gamma(\mu_\text{E}+\frac{\mu_\text{D}\alpha_\text{D}}{\alpha_\text{E}})\Gamma(\frac{z_{\text{E}}^{2}}{\alpha_\text{E}}+\frac{\mu_\text{D}\alpha_\text{D}}{\alpha_\text{E}})}{\Gamma(\mu_\text{D}+1)},
    % \end{align}
    \begin{align}
        \Phi = \frac{\Gamma(m_\text{E}-\frac{\mu_\text{D}\alpha_\text{D}}{\alpha_\text{E}})\Gamma(m_\text{D}+\mu_\text{D})\Gamma(\mu_\text{E}+\frac{\mu_\text{D}\alpha_\text{D}}{\alpha_\text{E}})}{\mu_\text{D}\left({z_{\text{D}}^{2}}/{\alpha_\text{D}}-\mu_\text{D}\right)\left({z_{\text{E}}^{2}}/{\alpha_\text{E}}+{\mu_\text{D}\alpha_\text{D}}/{\alpha_\text{E}}\right)},
    \end{align}
(ii) for $z_{\text{D}}^{2}/\alpha_{\text{D}}<\min\{\mu_{\text{D}},m_{\text{E}}\alpha_{\text{E}}/\alpha_{\text{D}}\}$,  
\begin{IEEEeqnarray}{lcl}
    \Phi = \frac{\Gamma(\mu_\text{D}-\frac{z_{\text{D}}^{2}}{\alpha_\text{D}})\Gamma(m_\text{E}-\frac{z_{\text{D}}^{2}}{\alpha_\text{E}})\Gamma(m_\text{D}+\frac{z_{\text{D}}^{2}}{\alpha_\text{D}})\Gamma(\mu_\text{E}+\frac{z_{\text{D}}^{2}}{\alpha_\text{E}})}{{z_{\text{D}}^{2}}\left(z_{\text{E}}^{2}+z_{\text{D}}^{2}\right)/({\alpha_\text{D}\alpha_\text{E}})},\IEEEeqnarraynumspace
\end{IEEEeqnarray}
and (iii) for $m_{\text{E}}\alpha_{\text{E}}/\alpha_{\text{D}}<\min\{\mu_{\text{D}},z_{\text{D}}^{2}/\alpha_{\text{D}}\}$,
\begin{align}
    \Phi =  \frac{\Gamma(\mu_\text{D}-\frac{m_\text{E}\alpha_\text{E}}{\alpha_\text{D}})\Gamma(m_\text{D}+\frac{m_\text{E}\alpha_\text{E}}{\alpha_\text{D}})\Gamma(\mu_\text{E}+m_\text{E})}{m_\text{E}\left({z_{\text{D}}^{2}}/{\alpha_{\text{D}}^{2}}-{m_\text{E}\alpha_\text{E}}/{\alpha_\text{D}}\right)\left({z_{\text{E}}^{2}}/{\alpha_\text{E}}+m_\text{E}\right)}.
\end{align}

\textit{Remark:} Making $\bar{\gamma}_{\text{d}}\rightarrow\infty$ while $\bar{\gamma}_\text{E}$ is fixed, it follows that the $\text{SOP}_{\mathcal{L}}^{\infty}$ can be written in the form $\text{SOP}_{\mathcal{L}}^{\infty}\approx\bar{\gamma}_{\text{D}}^{-\mathcal{G}_\text{d}}$, in which $\mathcal{G}_\text{d}$ denotes the secrecy diversity gain. 
From~(\ref{eq:SOPLOWEREXP_Asymptotic}), it follows that $\mathcal{G}_\text{d}=\alpha_{\text{D}}\Xi/2$.

\section{Numerical Results}\label{results}

This section presents results for secrecy metrics under $\alpha$-${\cal{F}}$ fading channels with pointing errors. 
The derived analytical results are validated through comparison with Monte Carlo simulations, demonstrating their accuracy. 
The simulations involve generating samples of the instantaneous SNRs for both the user and the eavesdropper, following the distribution in~\eqref{eq:PDFSNR}, and subsequently calculating the secrecy metrics. 
Without loss of generality, the path loss factor is normalized to $h_{\text{l}_i}=1$.

\begin{figure}[!t]
\begin{center}
       \includegraphics[width=0.89\linewidth]{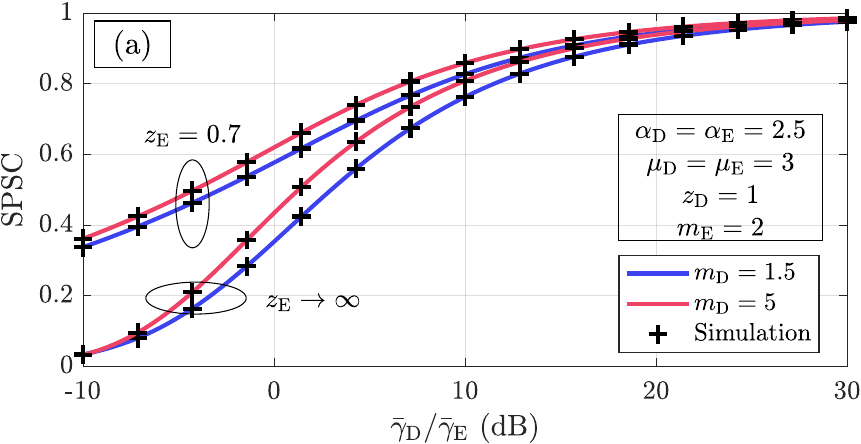}
       \includegraphics[width=0.89\linewidth]{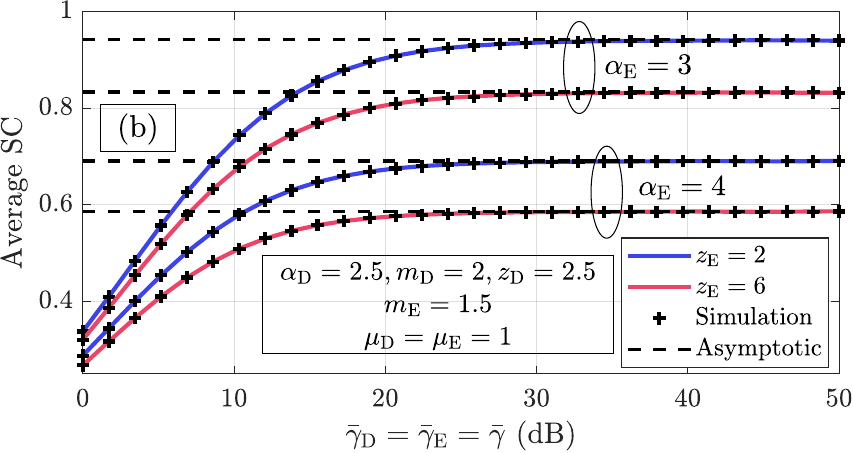}
       \includegraphics[width=0.89\linewidth]{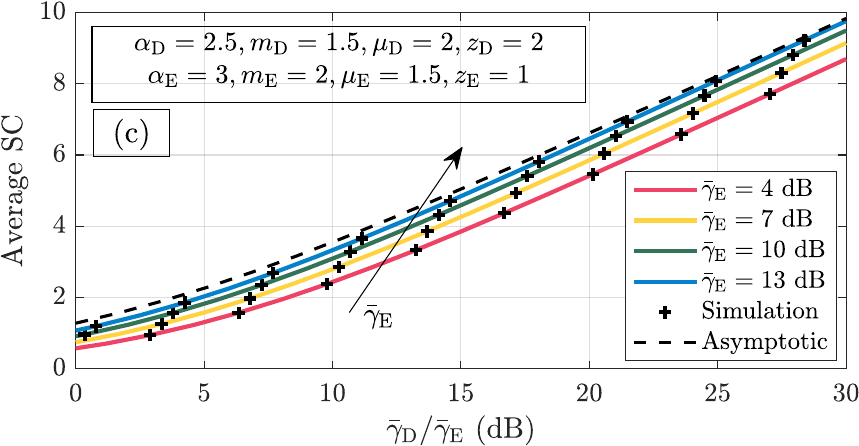}
       \includegraphics[width=0.89\linewidth]{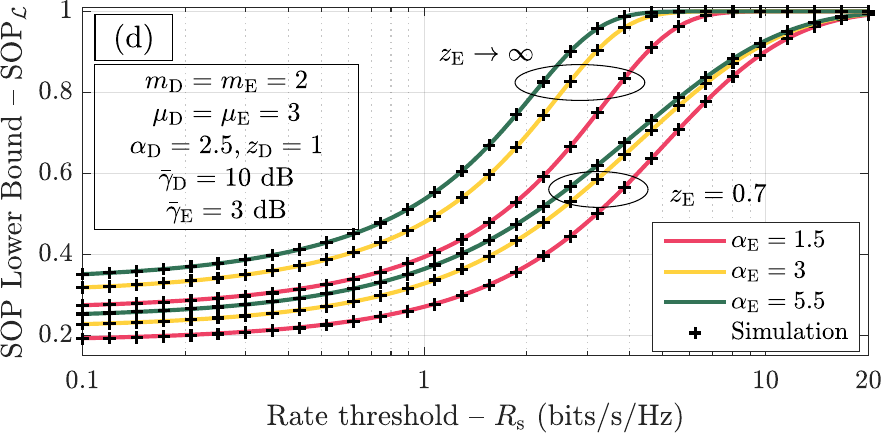}
       \includegraphics[width=0.89\linewidth]{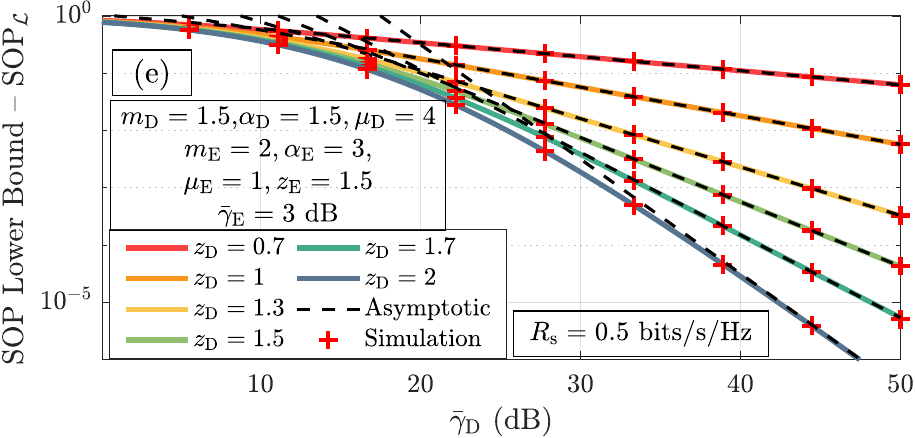}
   \caption{(a) SPSC and ASC curves versus (b) $\bar{\gamma}$ and (c)  versus $\bar{\gamma}_\text{D}/\bar{\gamma}_\text{E}$ under different scenarios. $ \text{SOP}_{\mathcal{L}}$ curves under different (d) $\alpha_\text{E}$ and (e) $z_\text{D}$ values.}
   \label{fig:1a}
\label{fig:1}
\end{center}
\end{figure}

Fig.~\ref{fig:1}(a) depicts SPSC curves as a function of the ratio $\bar{\gamma}_{\text{D}}/\bar{\gamma}_{\text{E}}$. 
Two scenarios are considered for the eavesdropper channel: the case with severe pointing errors ($z_{\text{E}} = 0.7$) and, as a benchmark, the case in the absence of pointing errors ($z_{\text{E}} \rightarrow \infty$). 
Furthermore, two values of $m_\text{D}$ are considered to evaluate the impact of the shadowing on the probability of SPSC. 
First, it is observed that as the $\bar{\gamma}_{\text{D}}/\bar{\gamma}_{\text{E}}$ ratio increases, there is a greater probability that the capacity corresponding to the legitimate user will exceed that of the eavesdropper, due to improved communication conditions. 
Conversely, when the eavesdropper has favorable channel conditions, corresponding to the scenario with $z_{\text{E}} \rightarrow \infty$, the probability of SPSC significantly decreases when the ratio $\bar{\gamma}_{\text{D}}/\bar{\gamma}_{\text{E}}$ is low. 
The impact of shadowing on the probability of SPSC is also evident, as a reduction in the mentioned metric occurs when the shadowing experienced by the user is more severe ($m_{\text{D}} = 1.5$).

Fig~\ref{fig:1a}(b) presents the exact and asymptotic ASC curves as a function of $\bar{\gamma}_\text{D}  = \bar{\gamma}_\text{E} = \bar{\gamma}$ under different pointing errors and nonlinearity conditions for the eavesdropper channel. 
For $\mu_\text{D}=\mu_\text{E}=1$, the shadowed Weibull fading model with pointing errors is obtained as a particular case of the study proposed in this work. 
To the best of the authors' knowledge, results for the shadowed Weibull model with the pointing error effect have not been presented in the literature concerning secrecy performance. 
As $\bar{\gamma}$ increases, the ASC gradually saturates at a given level.  
This result is expected, as enhancements in the eavesdropper’s link conditions diminish the user’s secrecy performance. 
Consequently, if $\bar{\gamma}_\text{D} = \bar{\gamma}_\text{E}$, increasing $\bar{\gamma}_\text{D}$ does not yield further improvements in secrecy performance beyond a certain limit. 
Higher values of $\alpha_{\text{E}}$ also promote better conditions for eavesdropper communications, thus reducing ASC. 
It can be noted that the ASC accurately converges to the value predicted by the asymptotic expression in \eqref{eq:AsympCapacity2}. 

Fig.~\ref{fig:1a}(c) presents the ASC curves as a function of the ratio $\bar{\gamma}_{\text{D}}/\bar{\gamma}_{\text{E}}$, with different fixed values for $\bar{\gamma}_{\text{E}}$. 
Since the eavesdropper’s conditions are held constant, the ASC increases as $\bar{\gamma}_{\text{D}}/\bar{\gamma}_{\text{E}}$ rises. 
It is also observed that, as $\bar{\gamma}_{\text{E}}$ increases, the corresponding curve approaches the asymptotic result predicted by~\eqref{eq:AsympCapacity2}, confirming the accuracy of this expression.

Fig.~\ref{fig:1a}(d) shows curves of $\text{SOP}_{\mathcal{L}}$ as a function of the rate threshold $R_\text{s}$, under different pointing errors and non-linearity conditions. 
The SOP increases with $R_\text{s}$, indicating that the combined channel and SNR conditions, along with the eavesdropper's influence, prevent achieving the desired secrecy rate. 
Similar to previous results, improvements in the eavesdropper's link conditions, such as the absence of pointing error and an increase in $\alpha_{\text{E}}$, deteriorate the secrecy performance.

Fig.~\ref{fig:1a}(e) shows curves of $\text{SOP}_{\mathcal{L}}$ as a function of  $\bar{\gamma}_{\text{D}}$, for different pointing errors conditions for the user and considering a rate threshold of $R_\text{s}=0.5$~{bits/s/Hz}.
The SOP improves as $\bar{\gamma}_{\text{D}}$ increases.
Note that lower $z_\text{D}$ values deteriorate the secrecy performance since the pointing error has a greater effect on the system.
Furthermore, observe that the asymptotic curves follow the lower bound of the SOP for a higher SNR regime.
As attested by our theoretical findings, the diversity order $\mathcal{G}_\text{d}$, represented by the slope of asymptotic curves, depended on the pointing error parameter $z_{\text{D}}$.

\section{Conclusions}\label{conclusao}

This paper has advanced the knowledge of PLS under $\alpha$-${\cal F}$ fading distribution with impairment of pointing errors, where the probability of SPSC, ASC, and the SOP secrecy metrics have been derived.
Curves have been presented for the mentioned metrics and validated using Monte Carlo simulations. 
A strong adherence between the theoretical and simulated curves has been noticed in all scenarios studied.

%\balance

\end{document}